\documentclass[twocolumn,preprintnumbers]{aastex63} 
\usepackage{afterpage}
\usepackage{capt-of}
\usepackage{diagbox}

\usepackage[figuresright]{rotating}
\usepackage{amsmath}
\usepackage{amssymb,amsmath,latexsym,graphics, graphicx,epsfig,multirow,comment,hyperref,appendix,feyn,slashed,xcolor,afterpage, makecell} 
\usepackage{booktabs}
\usepackage{tabularx}
\usepackage{wasysym}
\usepackage{placeins}

\newcommand {\be} {\begin {equation}}
\newcommand {\ee} {\end {equation}} 

\newcommand {\bes} {\begin {equation*}}
\newcommand {\ees} {\end {equation*}}

\newcolumntype{L}[1]{>{\raggedright\let\newline\\\arraybackslash\hspace{0pt}}m{#1}}
\newcolumntype{C}[1]{>{\centering\let\newline\\\arraybackslash\hspace{0pt}}m{#1}}
\newcolumntype{R}[1]{>{\raggedleft\let\newline\\\arraybackslash\hspace{0pt}}m{#1}}

\usepackage{csvsimple}

\newcommand{\Z}{\mathbb{Z}}
\newcommand{\N}{\mathbb{N}}
\newcommand{\R}{\mathbb{R}}
\newcommand{\C}{\mathbb{C}}

\makeatletter
\newcommand\footnoteref[1]{\protected@xdef\@thefnmark{\ref{#1}}\@footnotemark}
\makeatother

\DeclareRobustCommand{\Sec}[1]{Sec.~\ref{#1}}

\DeclareRobustCommand{\App}[1]{App.~\ref{#1}}
\DeclareRobustCommand{\Tab}[1]{Table~\ref{#1}}

\DeclareRobustCommand{\Fig}[1]{Fig.~\ref{#1}}

\DeclareRobustCommand{\Eq}[1]{Eq.~(\ref{#1})}

\renewcommand{\be}{\begin{equation}}
\renewcommand{\ee}{\end{equation}}

\newcommand{\V}[1]{{ \bf #1}}

\newcommand{\Gaia}{\textit{Gaia}~}
\newcommand{\vesc}{v_{\rm esc}}
\newcommand{\vmin}{v_{\rm min}}
\newcommand{\vobs}{v_{\rm obs}}
\newcommand{\Msun}{M_{\odot}}

\newcommand{\speed}{|\vec{v}|}

\begin{document}

\title{
Substructure at High Speed II: The Local Escape Velocity and Milky Way Mass with \Gaia DR2
}

\author{Lina Necib}
\affil{Walter Burke Institute for Theoretical Physics,
California Institute of Technology, Pasadena, CA 91125, USA}
\affil{Center for Cosmology, Department of Physics and Astronomy,
University of California, Irvine, CA 92697, USA}
\affil{Observatories of the Carnegie Institution for Science, 813 Santa Barbara St., Pasadena, CA 91101, USA}

\author{Tongyan Lin}
\affil{Department of Physics, University of California San Diego, La Jolla, CA 92093, USA}

\begin{abstract}
Measuring the escape velocity of the Milky Way is critical in obtaining the mass of the Milky Way, understanding the dark matter velocity distribution, and building the dark matter density profile. 
In \cite{methodology}, we introduced a strategy to robustly measure the escape velocity. 
Our approach takes into account the presence of kinematic substructures by modeling the tail of the stellar distribution with multiple components, including the stellar halo and the debris flow called the \Gaia Sausage (Enceladus). 
In doing so, we can test the robustness of the escape velocity measurement for different definitions of the ``tail" of the velocity distribution, and the consistency of the data with different underlying models. 
In this paper, we apply this method to the second data release of \emph{Gaia} and find that a model with at least two components is preferred.
Based on a fit with three bound components to account for the disk, relaxed halo, and the \Gaia Sausage, we find the escape velocity of the Milky Way at the solar position to be $\vesc = 484.6^{+17.8}_{-7.4}$~km/s.
Assuming a Navarro-Frenck-White dark matter profile, and taken in conjunction with a recent measurement of the circular velocity at the solar position of $v_c = 230 \pm 10$ km/s, we find a Milky Way concentration of $c_{200} = 13.8^{+6.0}_{-4.3}$ and a mass of $M_{200} = 7.0^{+1.9}_{-1.2} \times 10^{11} M_{\odot}$, which is considerably lighter than previous measurements.  \\
\end{abstract}

\section{Introduction} 
\label{sec:intro}

Since the initial discovery of dark matter (DM) \citep{1933AcHPh...6..110Z}, estimating the total mass and density profile of the Milky Way has been of crucial importance, providing a window into estimating the mass of the unseen DM. Various methods have been suggested to tackle this question, from modeling the density distributions of the different Galactic components \citep{1981ApJ...251...61C}, to the study of the fastest moving stars \citep{1982MNRAS.201..579A},  to fitting the local escape velocity of the stars as a way to constrain the local gravitational potential~\citep{1990ApJ}, to more complex methods that involve using large stream structures such as the Sagittarius stream to constrain the Milky Way potential at large distances \citep{2014MNRAS.445.3788G,2017ApJ...847...42D}. 

In recent years, a number of new phase-space structures have been discovered, which speaks to the success of hierarchical structure formation \citep{White:1977jf} and also suggests the need to re-examine methods to extract the Milky Way mass.
One of the many discoveries pioneered by \Gaia was the \Gaia Sausage or \Gaia Enceladus \citep{2018MNRAS.478..611B,2018Natur.563...85H}, which we will refer to as the Sausage in the remainder of this paper. The Sausage is the remnant of a merger that occurred 6 to 10 billion years ago between a galaxy with a stellar mass of $\sim 10^{8-9} M_{\odot}$ and the Milky Way~\citep{2018ApJ...863L..28M,2018ApJ...862L...1D,2018arXiv180704290L}.  This substructure is kinematically distinct from the stellar halo, with stars on extremely radial orbits \citep{2018ApJ...862L...1D}, and shifts the peak of the stellar speed distribution to lower values compared to the Standard Halo Model \citep{2019ApJ...874....3N}. 

In \cite{methodology}, we introduced a method to account for the presence of kinematic substructures in measurements of the escape velocity. Our work builds on the approach of  \cite{1990ApJ}, which modeled the tail of the stellar speed distribution as 
\be \label{eq:fv_tail}
f(v) \propto (\vesc - v)^k, \quad\quad \vmin < v < \vesc
\ee
where $\vesc$ is the escape velocity,  $k$ is the slope, and $\vmin$ is an arbitrary speed above which we define the ``tail" of the distribution. Many papers have used this formulation to infer the local escape velocity by fitting for the parameters $\vesc$ and $k$ with various datasets and assumptions \citep{2007MNRAS.379..755S,2014A&A...562A..91P,2018A&A...616L...9M,2019arXiv190102016D,2020arXiv200616283K}. These studies have found large correlations between $\vesc$ and $k$, and subsequently large errors on the escape velocity measurements. In order to reduce the error bars associated with $\vesc$, works such as \cite{1990ApJ,2007MNRAS.379..755S,2014A&A...562A..91P,2018A&A...616L...9M,2019arXiv190102016D} argue for imposing prior ranges on $k$ based on simulations. However, \cite{2019MNRAS.487L..72G} and \cite{methodology} showed that this approach can lead to underestimates of the escape velocity if the prior range on $k$ is too low, or overestimates of $\vesc$ if the prior range on $k$ is too high compared to the Milky Way.

In \cite{methodology}, hereafter Paper I, we developed a strategy for measuring $\vesc$ that does not rely on artificial prior ranges, and that accounts for the presence of substructure by including multiple bound components following \Eq{eq:fv_tail}. In particular, the choice of a rather low $\vmin = 300$ km/s is standard in the literature in order to increase statistics, but it is not clear that the speed distribution for $v > \vmin$ can be described by only one power-law component. For instance, in simulated halos with major mergers, the speed distribution can deviate significantly from the power-law form due to substructure~\citep{2019MNRAS.487L..72G}. Since it is known that the Sausage contributes a large fraction of non-disk stars in the solar neighborhood \citep{2019ApJ...874....3N,2019arXiv190707681N}, there is strong motivation to include multiple components in modeling the speed distribution.

To test the idea that $\vesc$ measurements might be biased by kinematic substructures, in Paper I we generated mock data that contains two components following \Eq{eq:fv_tail}, with a common $\vesc$ but different $k$. We attributed these to a ``relaxed" stellar halo component, which has a larger slope $k \sim$ 2-4, and to a Sausage component, with a lower slope $k_{S} =1$. These slopes are based on the analytic arguments of \cite{2019arXiv190102016D} for tracer populations with different levels of velocity anisotropy. When $\vmin$ is low, we found that a single-component fit to the mock data tends to overestimate $\vesc$ and also give larger error bars on $\vesc$, but as $\vmin$ is increased the single-component fit will converge to the correct value. A key conclusion of Paper I is that a robust inference of $\vesc$ requires testing fit dependence on $\vmin$ and on the number of components.

In this work, we use the methods developed in Paper I to estimate the escape velocity of the Milky Way at the solar position. We perform the analysis on the second data release of \Gaia (\Gaia DR2) subset that includes line-of-sight velocity measurements and passes the quality cuts of \Sec{sec:data}, as well as the subset of this sample with stars on retrograde motion. We test for the dependence of $\vesc$ on both $\vmin$ and number of components in order to ensure that the model selection is self-consistent and that results are robust to changes in data selection. 
This paper is organized as follows: we discuss the datasets used in \Sec{sec:data}, present the analysis and discuss best fit results in \Sec{sec:results}, and study implications for the mass of the Milky Way in \Sec{sec:mass_mw}.

\section{Data}
\label{sec:data}

We use \Gaia DR2 and focus on the subset of stars with radial velocity measurements \citep{2016A&A...595A...1G,2018arXiv180409365G} such that we can reconstruct the stellar speed. This subset is already a rather local sample of stars, but in order to restrict to a local measurement of the escape velocity, we implement a Galactocentric distance cut of $r_{\rm{GC}} \in [7.0, 9.0]$~kpc. 

We define the distance measurement as the inverse of the measured parallax ($1/ \varpi $). In order to make sure this assumption is valid, we also implement a cut on the parallax error, such that the fractional parallax error for each star is less than $10\%$. 
We use \cite{Bovy:2011aa} to analytically transform the proper motions and the radial velocities of \Gaia  into Galactocentric Cartesian coordinates, using the Local Standard of Rest $(U,V,W) = (11.1, 239.08, 7.25)$ km/s \citep{2010MNRAS.403.1829S}. We assume the position of the Sun in these coordinates is $(-8.12, 0, 0.02)$ kpc \citep{2019MNRAS.482.1417B}.

Of the stars passing the cuts above, the majority have fractional errors on the measured speed of $(\Delta v)/v < 5\%$ while a small number have fractional errors as large as 10\% (see \App{app:datasel}). In Paper I, we showed with mock data that parameter values could be robustly inferred with datasets where measurement errors are capped at 5\% or 20 km/s, but that results might become biased if errors become larger than 10\%. In this paper, we therefore place a cut of $(\Delta v)/v \le 5\%$  on the data sample to ensure the data is of comparable quality as the mocks tested in Paper I.  For the entire stellar sample where the fractional speed error is $\le$ 5\%, we find 1862 stars with measured speed greater than or equal 300 km/s. 
This sample size decreases to 161 stars for $v > 400$ km/s. 

In previous works such as \cite{2018A&A...616L...9M,2019arXiv190102016D}, only retrograde stars were modeled in order to avoid contamination from the disk. However, retrograde data also yields lower statistics, with only 442 stars above 300 km/s and 66 above 400 km/s, making it more difficult to draw strong conclusions from such a small sample.  Instead, with the analysis pipeline of Paper I, we can easily account for possible disk contamination by introducing an additional component in the model. We will therefore consider two sets of analyses: with the \Gaia DR2 subset as discussed above, where one of the components in our fit may be attributed to the disk, and with the subset of retrograde stars only. As we will see below, in fitting only the retrograde stars we are not able to obtain a robust estimate of $\vesc$. 

\begin{figure*}[t] 
   \centering
       \includegraphics[width=0.32\textwidth]{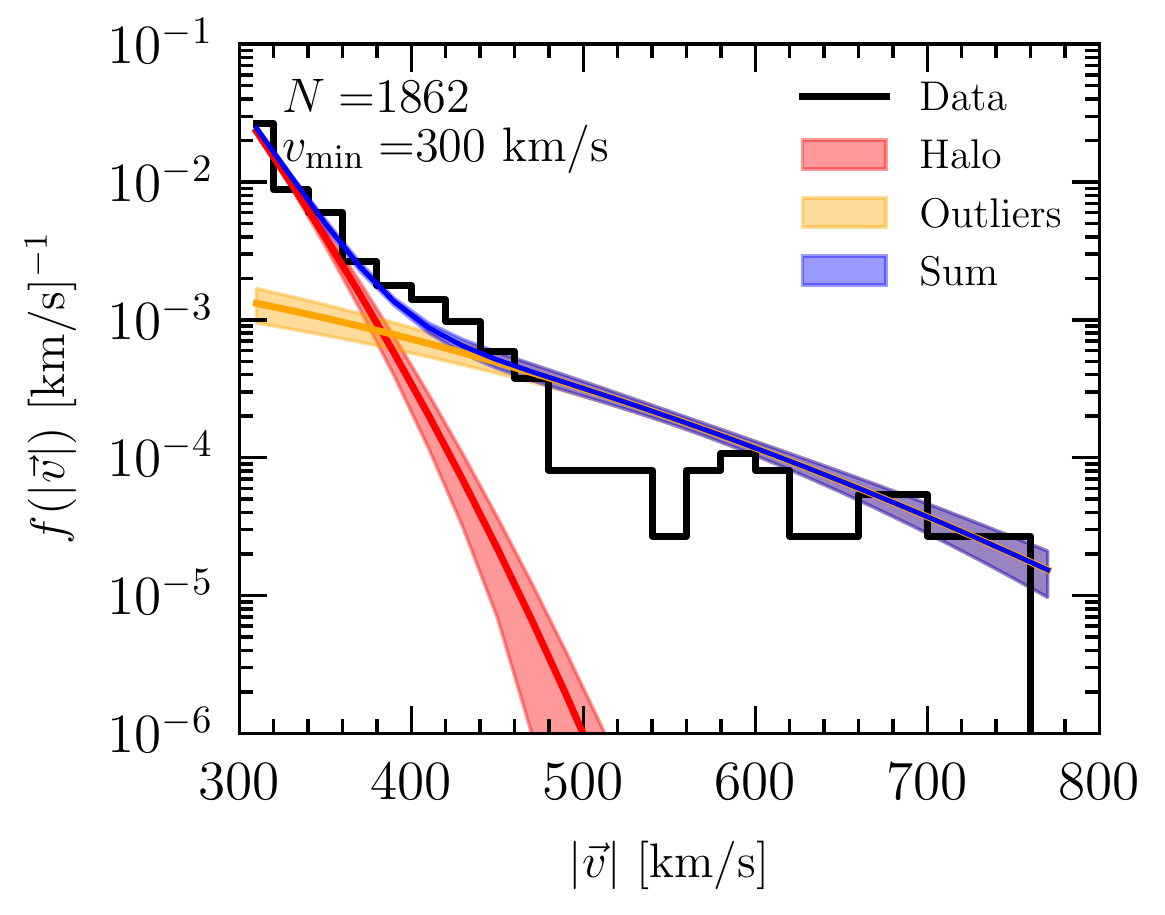} 
	\includegraphics[width=0.32\textwidth]{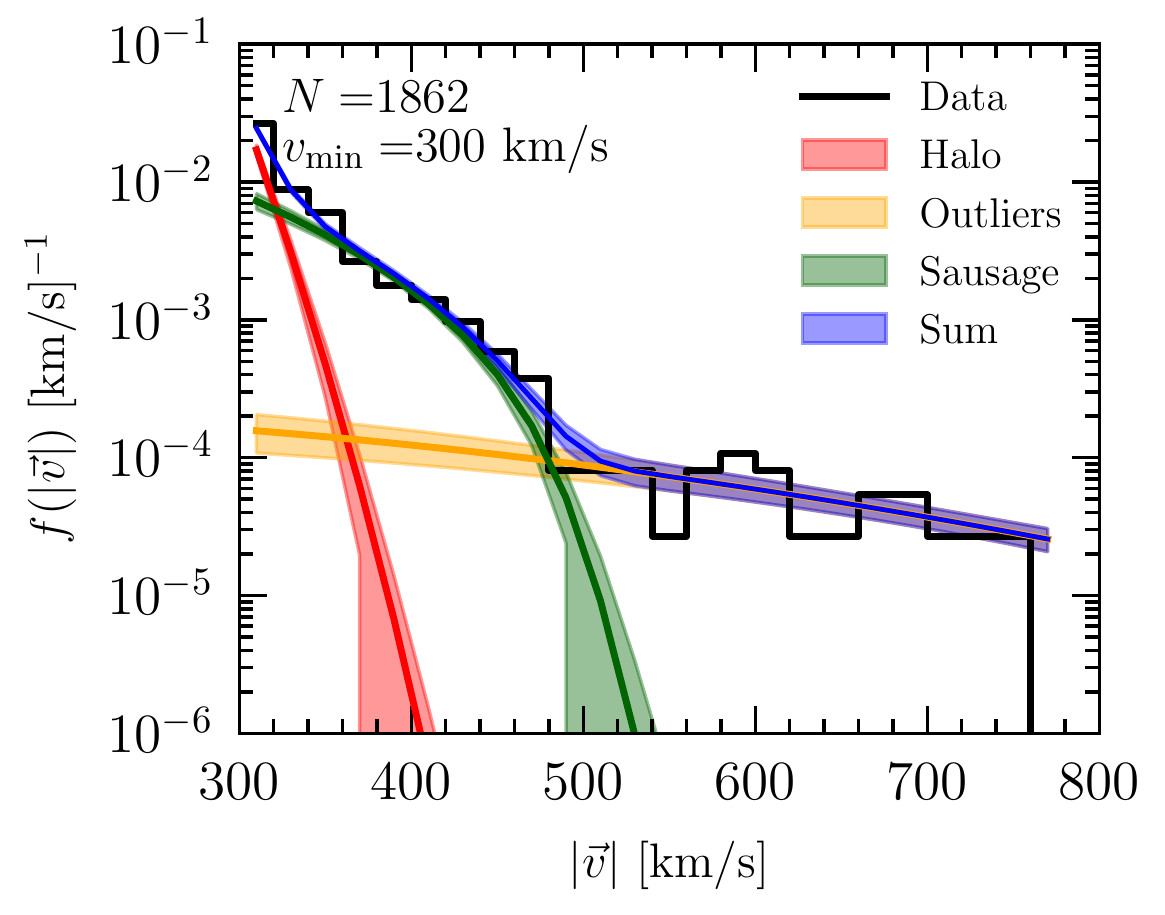} 
	\includegraphics[width=0.32\textwidth]{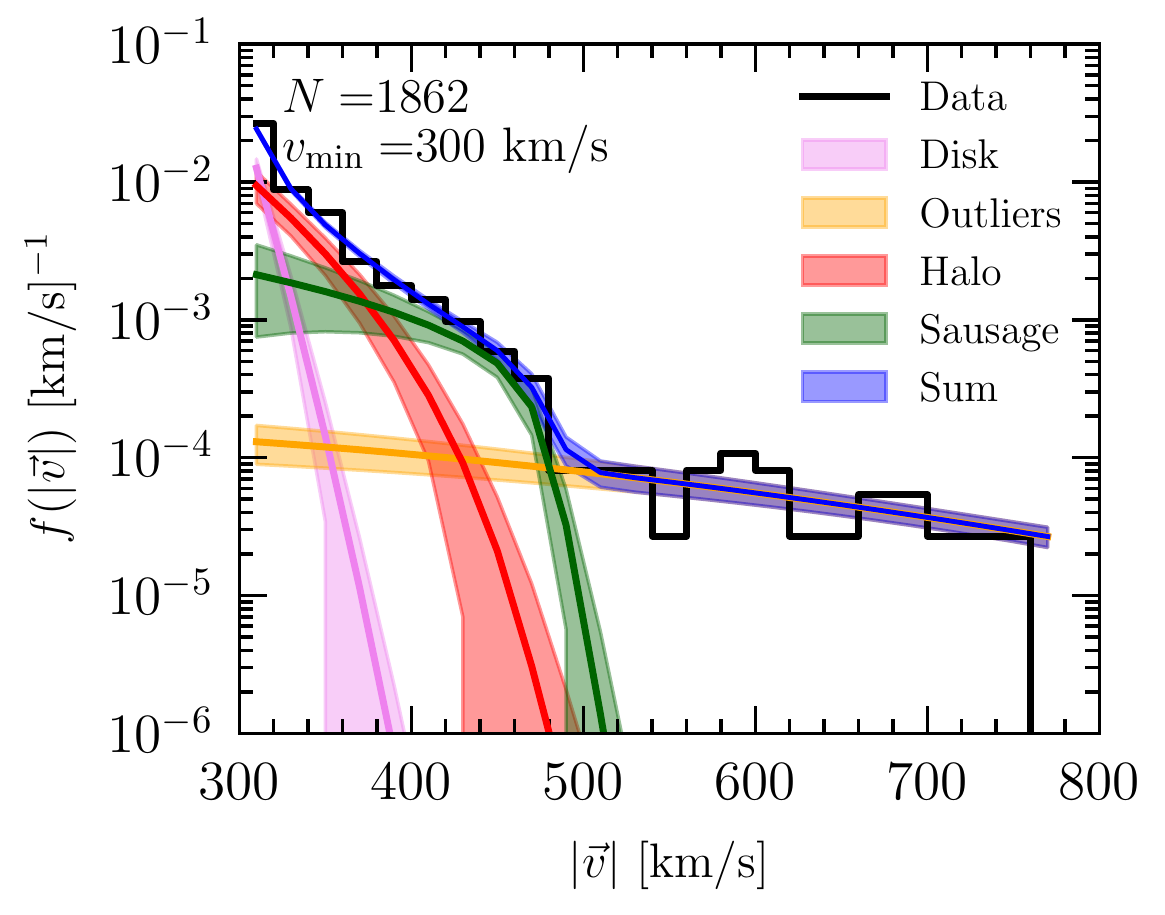} 
   \caption{Best fit results obtained from the full \Gaia data with one (left), two (middle), and three (right) bound components. The bound components are labeled with decreasing $k$ as Disk, Halo, and Sausage for convenience and in line with our expectation for how different components might behave, but the specific assignment cannot be determined from the fit. The solid lines are the best fit distributions, while the shaded regions are the 68\% containment regions obtained from the posteriors of the model parameters. The single-component model does not give a good fit to the data, while the results of the two- and three-component fits are largely consistent with each other and fit the data well. }
   \label{fig:comparison_results}
\end{figure*}

\section{Results}
\label{sec:results}

We first briefly summarize the analysis pipeline, while the complete details can be found in Paper I.
We model the stellar speed distribution above $\vmin$ with one, two, or three bound components, as well as an outlier component to account for unbound stars. Each of the bound components has a common $\vesc$ while the outlier distribution can extend above $\vesc$. Explicitly, the likelihood function for a single star is
\be
\mathcal{L} = f p_{\rm{out}}(\vobs | \sigma_{\rm{out}}) + (1 - f) \sum_{i} f_i p_i(\vobs| \vesc, k_i),
\ee
where $\vobs$ is the observed speed, $f_i$ the fractional contribution of the $i$-th component to the bound stars, and $p_i$ is the distribution of the $i$-th bound component. To obtain $p_{i}$, we consider \Eq{eq:fv_tail} with $\vesc$ and $k_i$ and convolve it with the measurement error for that star. The bound components are ordered such that $k_{i} > k_{i+1}$ when there are multiple components. To describe unbound outlier stars, we assume $p_{\rm{out}}$ is a Gaussian function with a dispersion of $\sigma_{\rm{out}}$ and that $f$ is the outlier fraction. Although the three component fit was not discussed in Paper I, it is a natural extension to include another component with the same $\vesc$ but a third slope $k_3$ and an associated fraction $f_3$. 
For the total likelihood, we sum over all stars in the dataset.

We used the Markov Chain Monte Carlo \textit{emcee} \citep{2013PASP..125..306F} to find the best fit parameters, using 200 walkers, 500 steps for the burnin stage, and 2000 steps for each run. We assume linear priors in $k_{i}$ and $f_{i}$ and log priors in $f, \sigma_{\rm out}$. We take a theory prior that is uniform in $1/\vesc$ in order to favor lower $\vesc$; note that for our main results, we find that the $\vesc$ posteriors are sharply peaked, such that this prior has a negligible effect on the results.

\subsection{All Data}

\begin{figure*}[t] 
   \centering
	\includegraphics[width=0.85\textwidth]{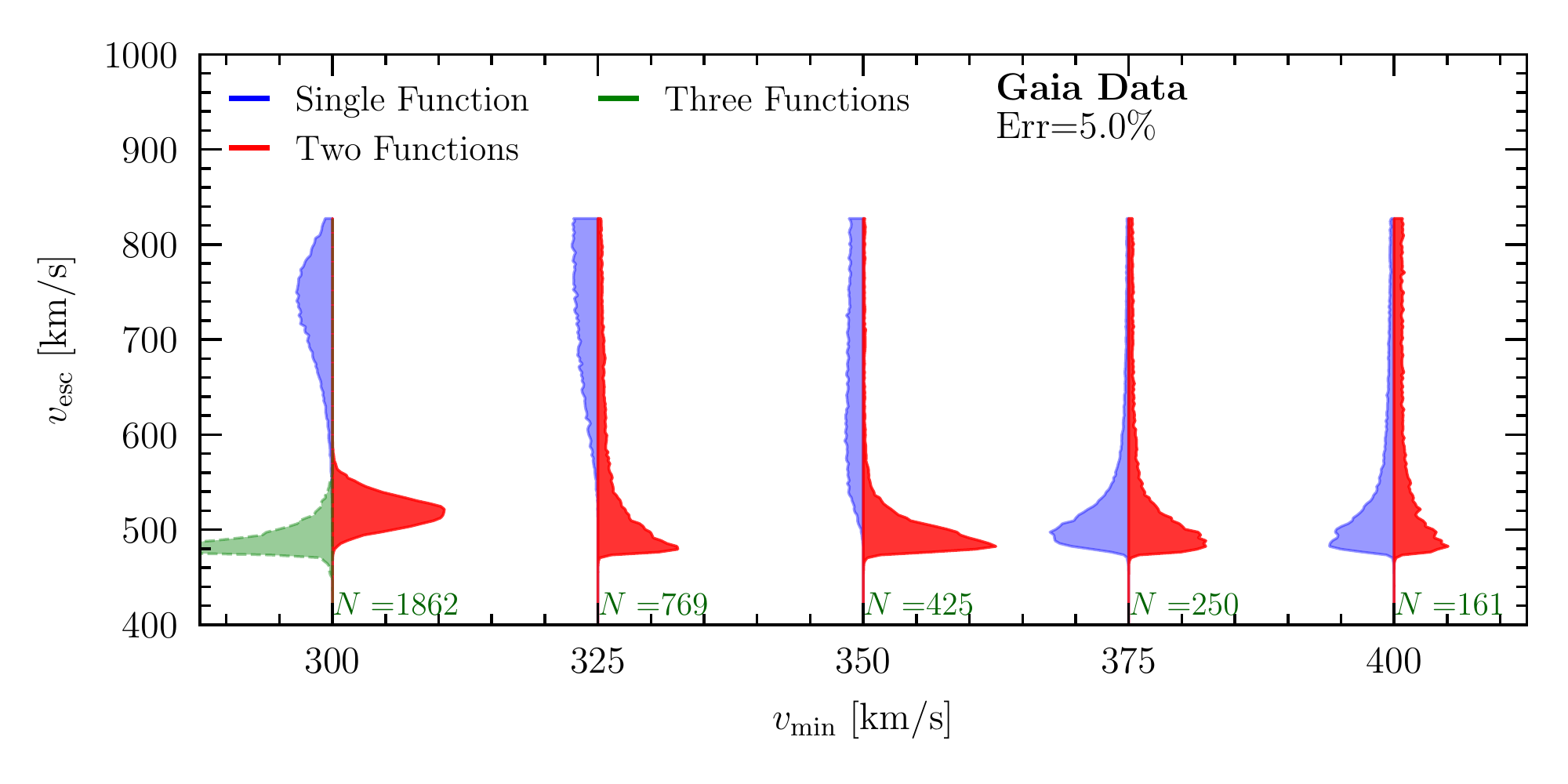} 	
   \caption{Posteriors in the escape velocity for different values of $\vmin$, from fitting the dataset satisfying quality cuts and fractional speed error $\le 5\%$. The posteriors for the single-component fits (shaded blue) drift with $\vmin$ for $\vmin < 375$~km/s, suggesting the need for an additional component in the fit. The posterior distributions for the two-component fit (shaded red) are much more stable with $\vmin$. For $\vmin = 300 $ km/s only, we show the posterior from a three-component fit (shaded green and dashed).  The labels $N$ indicate the number of stars in each sample.}
   \label{fig:violin}
\end{figure*}

We begin with the entire dataset satisfying the quality cuts of \Sec{sec:data} and with $\vmin = 300$ km/s. Because of the possibility of disk contamination, we consider up to three bound components. We will refer to the components in order of decreasing $k$ as the Disk, Halo, and Sausage for convenience, but note that in this analysis we cannot empirically determine a physical origin for each of the bound components. 

\Fig{fig:comparison_results} shows the best fit distributions obtained with one, two, or three bound components. First, we see that the single component fit (left panel) is not sufficient to describe the data. 
This is due to the fact that the fit for the bound component is anchored at the low end $|\vec{v}| \sim 300$ km/s, where there is a steep slope or high $k$. This leads to an underestimate of the number of stars at $v \sim 400$ km/s, which is partially compensated for by increasing the outlier fraction, ultimately leading to an overestimate of the number of stars at $\speed > 500$ km/s. In Paper I, we showed that when the underlying model contains multiple bound components, a single-component fit will bias $\vesc$ towards larger values.  Indeed, for this fit we found $\vesc = 735.9^{+50.5}_{-63.4}$~km/s.  

Meanwhile, we see that the two and three component models do provide good fits to the data and consistent results. The two function fit yields $\vesc = 519.6^{+17.4}_{-15.6}$~km/s and the three function fit yields $\vesc = 484.6^{+17.8}_{-7.4}$~km/s, which are consistent within two standard deviations. In both fits, we expect that the first component, or the highest $k$ component, will correspond to any disk contamination which drops steeply in $v$. Indeed, we find that the result for $k = k_{1}$ pushes up against our default prior of $k_{i} \in[0.1, 15]$ on the upper end; increasing the prior range to $k_{i} \in [0.1, 20]$, we still find a result that pushes at the edge of the prior. The resulting slope for this disk component is $k = 18.41^{+1.15}_{-1.94}$ and $k= 18.81^{+0.84}_{-1.45}$ for the two and three function fits, respectively. While it is expected that the disk distribution drops sharply for $\speed \gtrsim 300$ km/s, it might be concerning to have a fit pushing against the prior here. Therefore, we next turn to the dependence of our results on $\vmin$. Studying the fit dependence with increasing $\vmin$ will provide additional tests of robustness as well as  further eliminate the disk contamination. 

Our analysis is repeated for $\vmin \in [300, 235, 350, 375, 400]$ km/s. As shown in Paper I, this provides a consistency test for the underlying model; a $\vesc$ posterior which drifts with $\vmin$ suggests that the model is missing some important component.
In \Fig{fig:violin}, we show the posterior distribution of the escape velocity for the single and two component fits for different values of $\vmin$. As $\vmin$ increases, $\vesc$ in the single-component fit drifts towards lower values, converging towards similar results for $\vmin = 375$ km/s and $\vmin = 400$ km/s. This indicates that additional components are needed to describe the data below $\speed \sim 375$ km/s, but that it is dominated by a single component for $\vmin \ge 375$ km/s.  Indeed, the posterior distributions for the two-component fit are much more stable with $\vmin$ and consistent with the single component fit for $\vmin \ge 375$ km/s. However, for $\vmin = 300$ km/s the 2-component posterior is peaked at higher $\vesc$ values, suggesting that a third component might be present. Thus, we also show the posterior for the 3-component fit and $\vmin = 300$ km/s. This result is consistent with the posteriors for single and two-component fits at larger $\vmin$. (We do not show the posteriors for the 3-component fit and $\vmin \ge 325$ km/s because those are largely unconstrained, indicating that we have too many degenerate parameters.)

\begin{figure}[t] 
   \centering
	\includegraphics[width=0.48\textwidth]{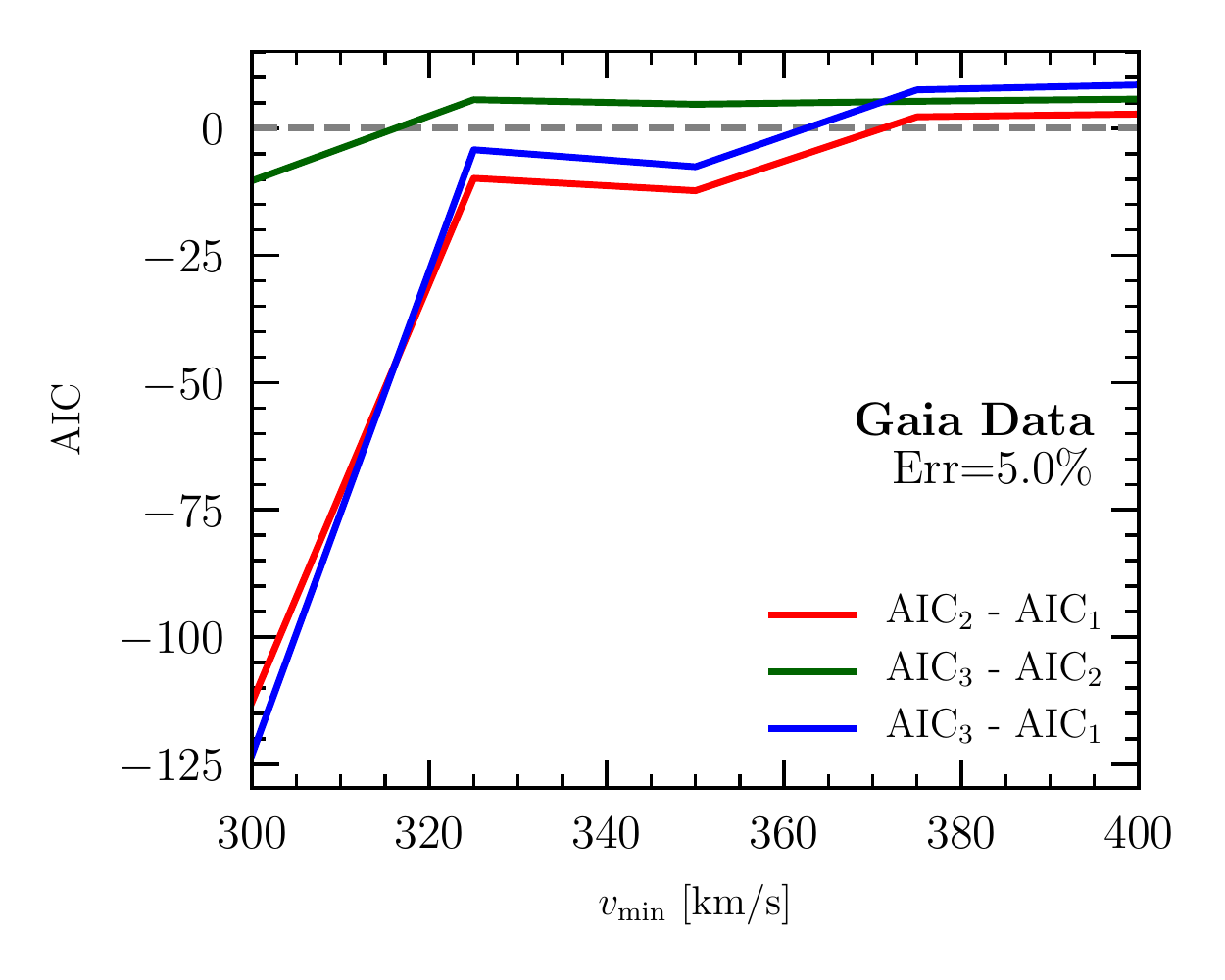} 
   \caption{We use the $\Delta$AIC to compare goodness-of-fit between models with different numbers of bound components. The plot shows results using the full \Gaia dataset, and the subscript indicates number of components in the fit. When $\Delta$AIC is negative, the first model is preferred.  }
   \label{fig:delta_aic}
\end{figure}

In order to compare goodness-of-fit with different numbers of bound components, we calculate the AIC \citep{aic} as discussed in Paper I. The AIC is defined as AIC = $2 s - 2 \log(\hat {\cal L})$ where $s$ is the number of model parameters and  $\log(\hat {\cal L})$ is the maximum log likelihood. To compare models, we compute the difference $\Delta \rm{AIC}_{ab} = \rm{AIC}_a - \rm{AIC}_b$ between fits with $a$ or $b$ bound components, where the model with the lowest AIC is preferred. We show the resulting $\Delta$AIC in \Fig{fig:delta_aic}, where we compare the model with one, two, and three components. For $\vmin = 300$ km/s, where we expect a large disk contamination, we find that the two and three component models are strongly preferred over a single function case. This is visible from the first panel of \Fig{fig:comparison_results}, where a single bound component clearly does not describe the data as well. As expected however, the single function is preferred for $\vmin \ge 375$ km/s. This is because at high $\speed$ the distribution is dominated by a single distribution, and extra model parameters are penalized. It is interesting to see that the three function fit is marginally better than the two function fit only for $\vmin = 300$ km/s, beyond which it leads to an overfit. This implies that the data is determined by three components for $\vmin = 300$ km/s, but that the disk is largely subdominant past $\vmin = 325$ km/s.

From this discussion, we see a consistent story emerge: there is a sharply peaked component of stars near $\speed \sim 300$ km/s, along with two other bound components that dominate at higher speeds.  A single component dominates the tail for $\speed > 375$ km/s. These could already be seen in the three-component fit shown in \Fig{fig:comparison_results}, but is further supported by studying fit dependence with $\vmin$ and the $\Delta$AIC. We conclude that a single-component model is not a self-consistent description of the data for low $\vmin$. 
Our analysis allows us to identify the ``tails'' where a single or two-component fit is actually robust. In \Sec{sec:fitsummary}, we will summarize the key results for $\vesc$ and also compare with previous studies. 

It is important to note that our framework only shows that this data is better modeled by multiple components with power law distributions. It cannot show that each component has a distinct physical origin. Doing so requires further studies. Nevertheless, the behavior of the different components appears to be consistent with contributions from the Disk, relaxed Halo component, and Sausage, and here we summarize the fit results for those components. Additional results can be found in \App{app:corner}.

As mentioned in Paper I, we do not have \emph{a priori} information about which component is the Halo and which is the Sausage; the labels of \Fig{fig:comparison_results} are based on the assumption that the Sausage has the lowest slope $k$. Working with that assumption, we find that the posterior on the slope of the Sausage is $k_{S} =  0.88^{+0.81}_{-0.49}$  from the three function fit at $\vmin = 300$ km/s. This is similar to the range of slopes in simulations with Sausage-like mergers, $k_{S} \in [1,2.5]$, as obtained by \cite{2019arXiv190102016D}. In the same fit, the fractional contributions of the Halo and Sausage are $0.43^{+0.08}_{-0.15}$ and $0.22^{+0.13}_{-0.08}$, respectively. This leads to the Sausage component being  $0.34^{+0.20}_{-0.12}$ of these non-disk stars for $\speed > 300$ km/s. Using the values of the best fit slopes, we find that the Sausage component rises to $0.59^{+0.21}_{-0.17}$ of non-disk stars for $\speed > 350$ km/s.  This is consistent with the Sausage fraction obtained in the two-component fit with $\vmin = 350$ km/s, $f_S = 0.75^{+0.07}_{-0.13}$. These fractional contributions are important in understanding the composition of the DM distribution and impacts predicted signals in DM searches, which was studied in \cite{2018arXiv181012301N,2019arXiv190707190N}.

\subsection{Retrograde Data} 

\begin{figure}[t] 
   \centering
	\includegraphics[width=0.48\textwidth]{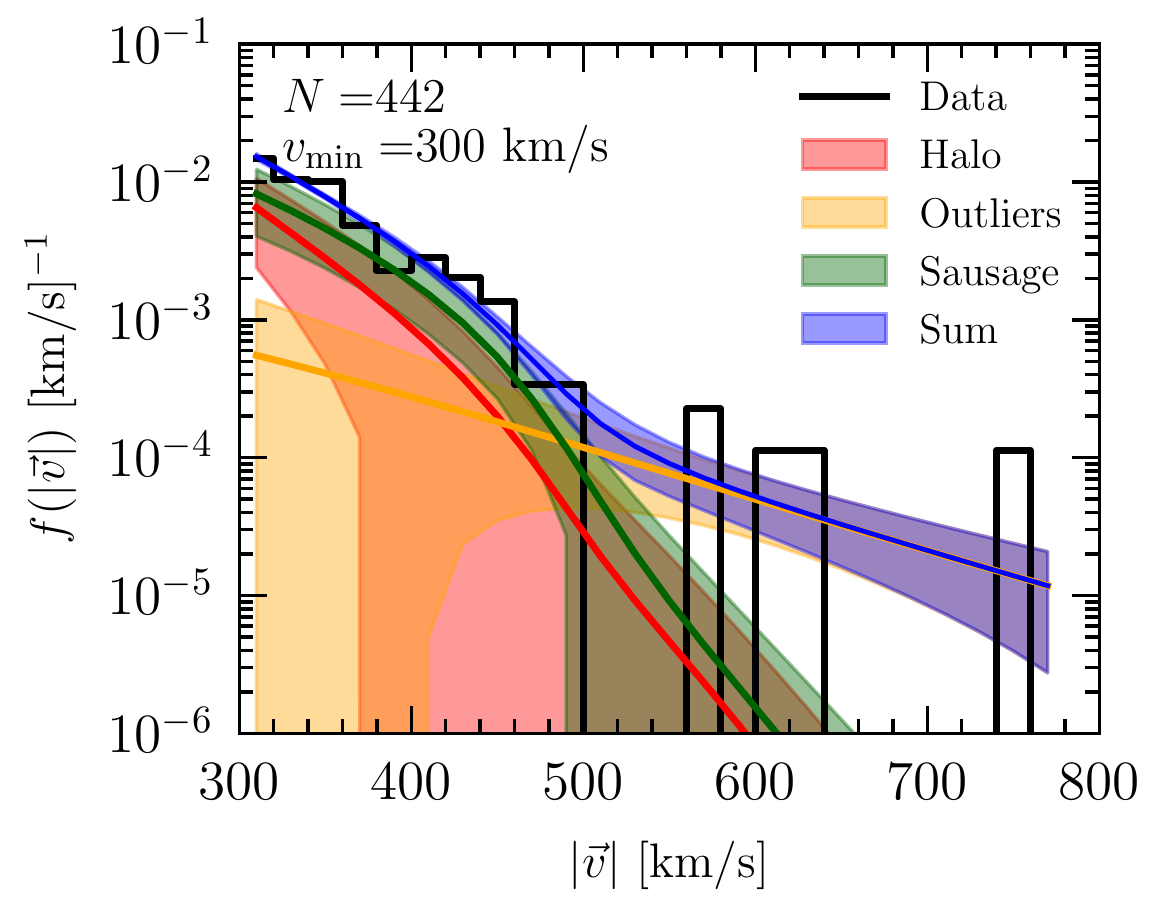} 
   \caption{Two function fit only to the retrograde stars satisfying quality cuts and with fractional speed errors capped at 5\%. The bound components and outlier population are much more poorly constrained than the fits to all stars, and we do not obtain a robust $\vesc$ result. }
   \label{fig:retro}
\end{figure}

We next perform the same analysis with retrograde stars only. Previous studies have applied this cut on the data to avoid disk contamination. Similarly, here we consider a fit with two bound components since we do not expect any disk contribution for $\vmin = 300$ km/s. We show the resulting fit in \Fig{fig:retro}, where we see that the 68\% containment region (shaded bands) for all the components is much larger compared to the fit to all stars in \Fig{fig:comparison_results}. It is expected that the uncertainties on the components should be larger given that the statistics is about 2.5 times lower, but the spread in the bands here reflects more than that. Focusing on the tail of the bound components, we see that there is significantly more spread in $\vesc$ and in fact the fit gives $\vesc = 710.3^{+216.5}_{-177.9}$ km/s.

\begin{figure}[t] 
   \centering
	\includegraphics[width=0.48\textwidth]{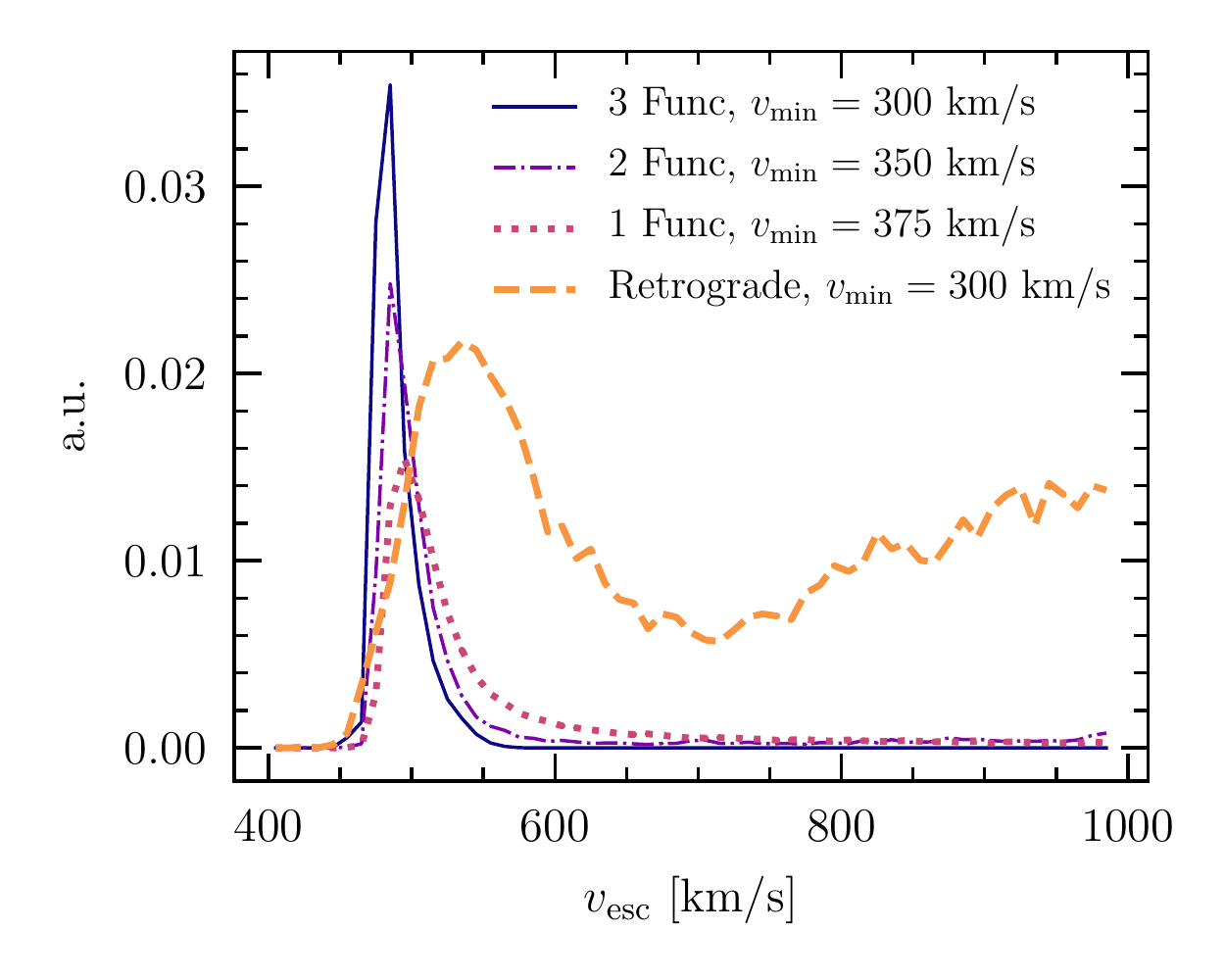} 
   \caption{Posteriors for $\vesc$ from fits to all data with 1, 2, or 3 bound components, and from a fit to retrograde data with 2 bound components. Full data posterior distributions have been normalized to unity, while the retrograde data has been multiplied by a factor of 6 to clarify the non-convergence of the posterior. For each of fits to all data, we selected the lowest $\vmin$ valid for that number of components. For the retrograde data, we do not obtain a convergent result. }
   \label{fig:vescposterior}
\end{figure}

The reason for the much larger uncertainty in $\vesc$ here is due to the difficulty in constraining the outlier population. In particular, with the dataset in \Fig{fig:retro}, it is difficult to distinguish whether the stars with speeds of 500-600 km/s are likely to come from the outlier component or the bound components. We can further understand this outlier confusion by looking at the escape velocity posterior, which is shown in \Fig{fig:vescposterior}. Here we see that the posterior for the 2-component fit to retrograde stars is actually double-peaked. One peak is at low $\vesc$, at values roughly consistent with fits to all stars. Meanwhile, the peak at high $\vesc$ is correlated with a lower outlier fraction, where those high speed stars near $\speed \sim 600$ km/s can also partly be modeled with the bound components. 

As a result, taking $\vesc = 710.3^{+216.5}_{-177.9}$ km/s for retrograde stars is not a robust result. In \App{app:retrograde}, we show the fit to retrograde stars for different $\vmin$ and see similar results with non-convergent fit results and double-peaked posteriors. This is expected since the outlier confusion cannot be removed with larger $\vmin$. In fact, the larger number of high-speed stars in the full data sample helps pin down the outlier component and thus also $\vesc$ of the bound components. With limited statistics in the retrograde sample, modeling the tail of a distribution is particularly sensitive to shot noise in the outliers. Furthermore, applying a cut for retrograde motion might be shaping the sample such that it is not kinematically complete. The possible presence of additional kinematic structures was studied in Paper I by the introduction of small peaked structures, and we found that it could strongly bias the results. Given these factors, we will consider the fit to all of the data as our main result.

\subsection{Comparison with previous results \label{sec:fitsummary} }

\begin{table}[t]
\begin{center}
\begin{tabular}{|l| c | c | c | c |}
\hline 
Dataset & Functions & $\vmin$ & $\vesc$  & $M_{\rm{200, \rm{tot}}} $ \\
 &  & [km/s] & [km/s] &  $[10^{11} M_{\odot}$] \\
 \hline 
 \hline 
All data & 3 & 300 & $484.6^{+17.8}_{-7.4}$ & $7.0 ^{+1.9}_{-1.2}$ \\
All data & 2 & 350 & $498.1^{+63.9}_{-16.0}$ & $7.7 ^{+2.6}_{-1.5}$ \\
All data & 1 & 375 & $515.1^{+113.3}_{-25.0}$ & $8.8 ^{+3.9}_{-2.0}$ \\
 \hline
\end{tabular}
\end{center}
\caption{\label{tab:results} Best fit values of the escape velocity at the solar position, as well as the total mass of the Milky Way obtained through the assumptions of \Sec{sec:mass_mw}. }
\end{table}

The main results for $\vesc$ are summarized in \Fig{fig:vescposterior} and \Tab{tab:results}. For these fits, we selected the lowest $\vmin$ that was consistent with the number of bound components in the fit. This was determined by the lowest $\vmin \in [300, 325,350,375,400]$ km/s  that yielded convergent and stable results. From \Fig{fig:vescposterior} and \Tab{tab:results}, we see that the results for 1, 2, or 3 bound components are also all consistent with each other.

Previous studies all used a single function of the form of \Eq{eq:fv_tail}. To deal with a large degeneracy in $\vesc$ and $k$, they typically impose an artificial prior on the slope $k$. \cite{2014A&A...562A..91P,2018A&A...616L...9M} adopt a prior of $k \in [2.3, 3.7]$, obtained using cosmological simulations and finding the slopes of the tail of the halo stars in their Milky Way realizations. Meanwhile, \cite{2019arXiv190102016D} used a lower prior of $k \in [1,2.5]$ given that cosmological simulations with merger events similar to the \Gaia Sausage had slopes in that range. These priors strongly affect the measured escape velocity, giving rise to nonconvergent results as can be seen in the corner plot of \cite{2019arXiv190102016D}. 

As discussed in Paper I, the use of a single function where the fit requires two leads to an averaging of the slopes $k$. 
For example, in $\vmin = 350$ km/s where the two function fit suffices, we find two components with $k = 12.72^{+4.09}_{-3.92}$ and $k_S = 1.34^{+1.29}_{-0.51}$, while a single function fit yields $k = 8.13^{+4.45}_{-3.68}$. One can see in this case that although $k_S$ is indeed within the range suggested by \cite{2019arXiv190102016D}, using a single function fit and a limited prior in $k$ can artificially inflate the escape velocity value.  

Given this, it is not surprising that our results for $\vesc$ are lower than previous studies. \cite{2019arXiv190102016D} finds $\vesc = 528^{+24}_{-25}$ km/s with a prior range of $k \in [1,2.5]$, while \cite{2018A&A...616L...9M} finds $\vesc =580 \pm 63 $ km/s with a prior range of $k \in [2.3,3.7]$. Our results are consistent with a recent study by  \cite{2020arXiv200616283K}. From an analysis including radial velocities and without imposing a prior in $k$, they find  $\vesc = 497^{+53}_{-20}$~km/s where the quoted errors are the 99\% confidence intervals. Using a larger sample of stars with only tangential velocities, they report a lower limit of $\vesc = 497^{+40}_{-24}$~km/s at the solar radius, again giving the 99\% confidence interval.

\section{Mass of the Milky Way}
\label{sec:mass_mw}

The escape velocity at a certain radius is related to the gravitational potential $\Phi$. Therefore, measurement of the escape velocity can translate to a measurement of the mass of the Milky Way, once profiles of the different mass components are assumed. For an isolated halo $\vesc (r) = \sqrt{2 |\Phi(r)|}$, but in practice we must select a limiting radius beyond which stars can become unbound. A more realistic assumption is that stars are bound within a few times $r_{200}$, where $r_{200}$ is the radius at which the galaxy's mass is 200 times the critical mass of the universe. For ease of comparison, we will adopt the same definition as \cite{2019arXiv190102016D} by taking the limiting radius as $2 r_{200}$, with $\vesc(r_{\odot}) = \sqrt{2 |\Phi (r_{\odot}) - \Phi(2r_{200}) |}$.  

To recover the mass of the Milky Way, we make the following assumptions on the different baryonic and DM components, similarly to \cite{2019arXiv190102016D}, which assumes model I of \cite{2017A&A...598A..66P}:
\begin{itemize}
\item The bulge is modeled as a Plummer profile \citep{1911MNRAS..71..460P} of mass $M_{\rm{bulge}} = 1.067\times10^{10} \Msun$ and a scale radius $b = 0.3$ kpc.
\item The thin disk is modeled as a Miyamoto-Nagai profile \citep{1975PASJ...27..533M} of mass $M_{\rm{thin~disk}} = 3.944\times10^{10} \Msun$, a scale radius $r_{\rm{thin~disk}} = 5.3$ kpc, and a height radius of $z_{\rm{thin~disk}} = 0.25$ kpc. 
\item The thick disk is modeled as Miyamoto-Nagai profile of mass $M_{\rm{thick~disk}} = 3.944\times10^{10} \Msun$, a scale radius $r_{\rm{thick~disk}} = 2.6$ kpc, and a height radius of $z_{\rm{thick~disk}} = 0.8$ kpc. 
\item The Dark Matter profile is modeled as an Navarro-Frenk-White (NFW) \citep{1996ApJ...462..563N} profile of mass $M_{200}$ and concentration parameter $c_{200}$, which we will fit for. We take the Hubble constant $H = 70$ km s$^{-1}$ Mpc$^{-1}$, the matter abundance $\Omega_M = 0.3$ \citep{Ade:2015xua}, and the overdensity taken with respect to the critical mass of the universe.
\end{itemize}

\begin{figure}[t] 
   \centering
	\includegraphics[width=0.43\textwidth]{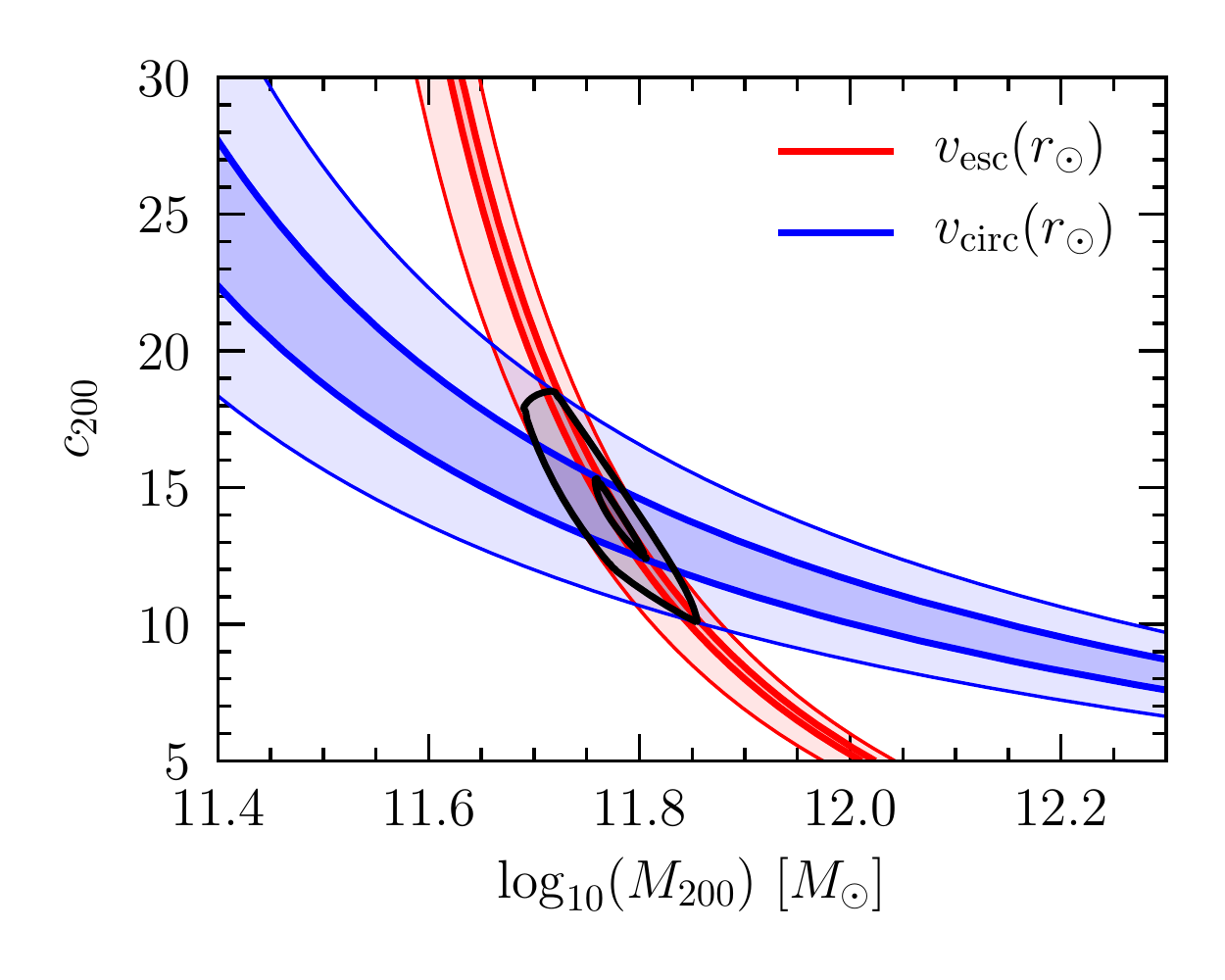} 
   \caption{Mass and concentration of the DM halo, constrained by the circular velocity measurement (blue) from \cite{2019ApJ...871..120E}, and by the escape velocity (red) studied in this work. For $\vesc$ we take the posterior from the three-function fit and $\vmin=300$ km/s. The contours are 68\% and 95\% CL, and the black lines show the combined constraints. }
   \label{fig:mass_concentration}
\end{figure}

To translate the posterior distribution of $\vesc$, marginalized over all other parameters, into a posterior in the enclosed mass-concentration $M_{200} - c_{200}$ space, we use \textsc{galpy} \citep{2015ApJS..216...29B} to compute the escape velocity of the summed potentials assumed above in a grid of $M_{200} $ and $c_{200}$. We then plot the PDF of each point in $M_{200} $ and $c_{200}$ using the interpolated version of the escape velocity posterior distribution, and show in \Fig{fig:mass_concentration} the 68\% and 95\% containment regions based on $\vesc$. Similarly, we overlay constraints based on the circular velocity $v_{\rm{circ}} = 230{\pm 10}$~km/s from \cite{2019ApJ...871..120E}. This is important as the escape velocity gives information on the mass of the Milky Way at large distances, while the circular velocity constrains it within the solar circle. 

\begin{figure}[t] 
   \centering
	\includegraphics[width=0.45\textwidth]{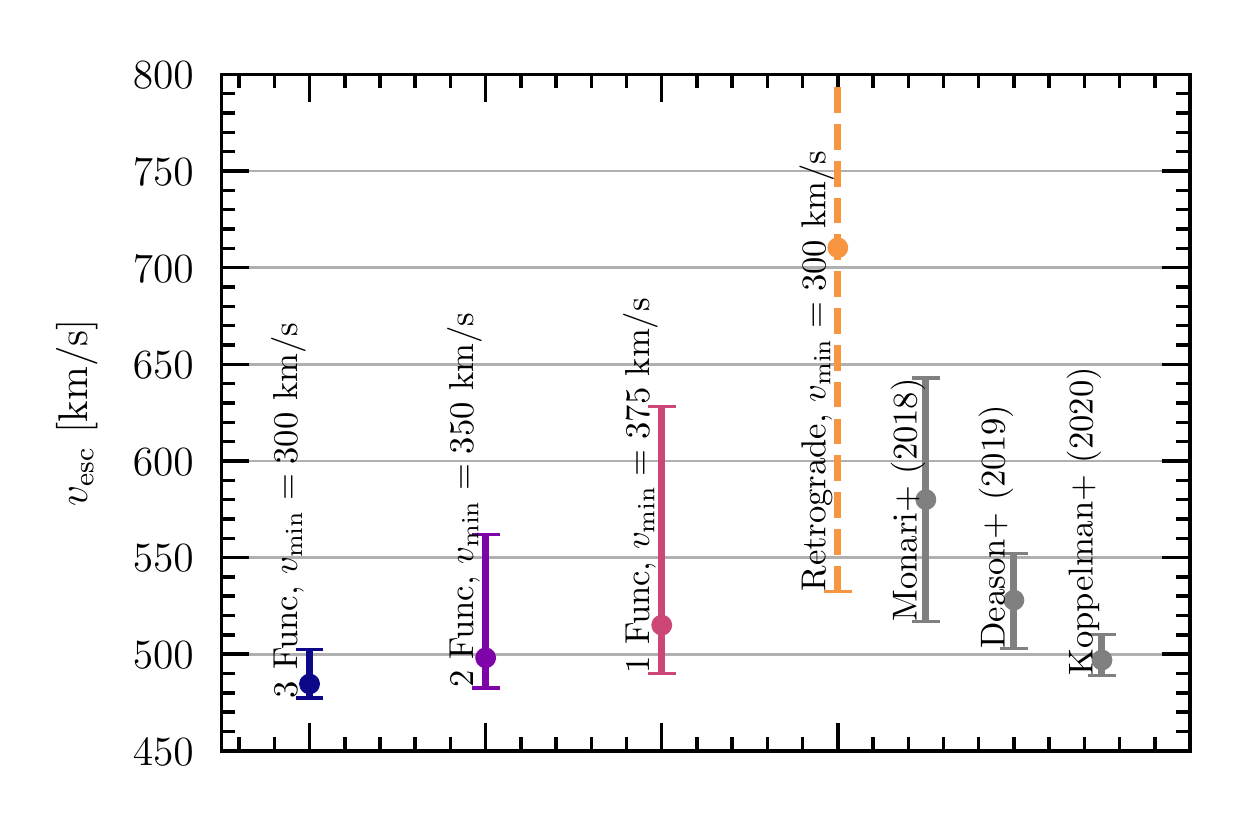} 
	\includegraphics[width=0.45\textwidth]{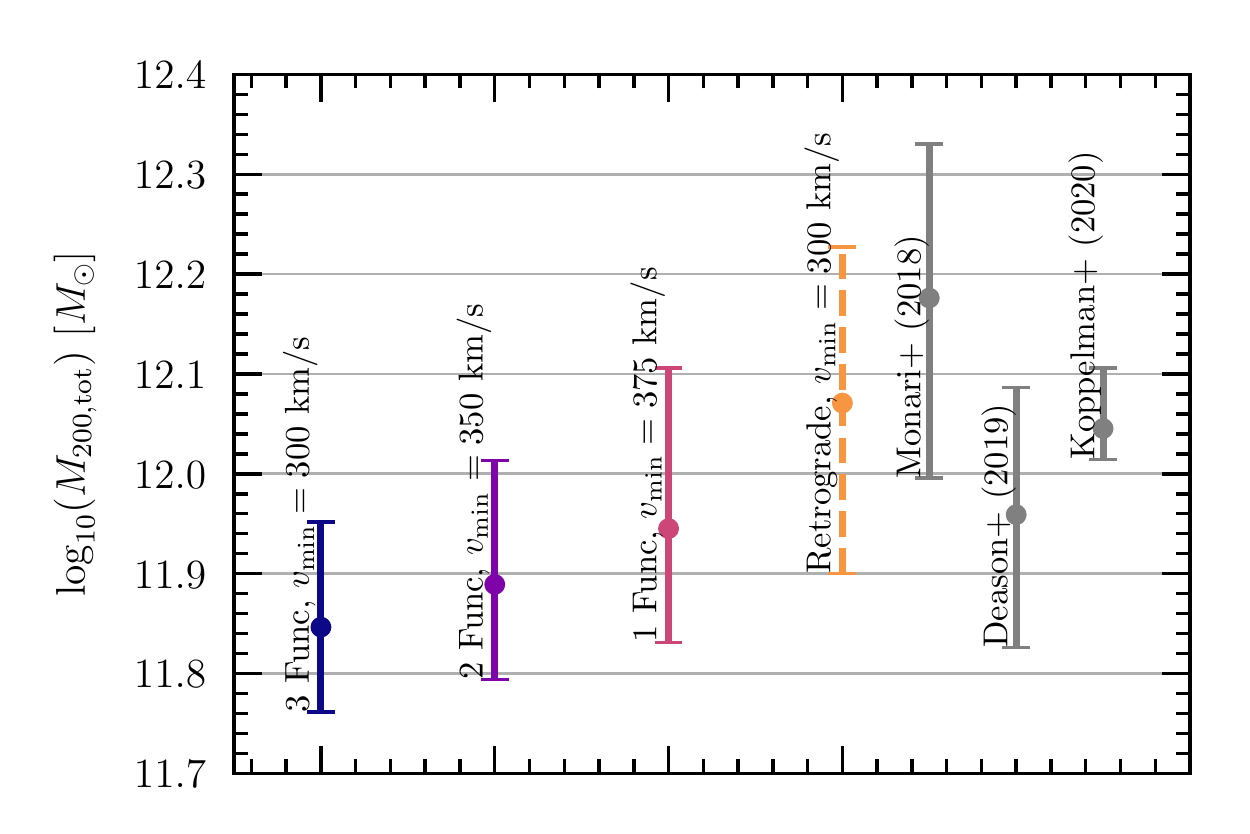} 
   \caption{Best fit values of the escape velocity (top) and total mass (bottom) of the Milky Way for the different analyses discussed in \Sec{sec:results}. The analyses of the full data is the most robust, and tends to be lower than previous measurements; it is within a single standard deviation of \cite{2019arXiv190102016D} and \cite{2020arXiv200616283K} and lower than measurements by \cite{2018A&A...616L...9M}. 
   The result based on the retrograde data is shown for completeness, but that analysis did not give a robust $\vesc$ (see \Fig{fig:retro}). We show the escape velocity measurement of \cite{2020arXiv200616283K}  where we divide their errors by 3, as they quote 99\% confidence intervals while we quote 68\%. For the mass measurement, \cite{2020arXiv200616283K} multiply the escape velocity estimate by 10\%, motivated by the findings of \citep{2019MNRAS.487L..72G} for potential biases. The mass plotted above is the one corrected with this factor. }
   \label{fig:mass_results}
\end{figure}

We now compare our findings to the literature. We define the total mass of the Milky Way as the mass contained within $r_{200}$ of the DM halo, in addition to the baryons, including the bulge, thin, and thick disk as described above.  In \Fig{fig:mass_results}, we show the results for the \textit{total} mass based on the four analyses summarized in \Sec{sec:fitsummary}. For analyses with higher $\vmin$, the errors on the escape velocity increases, as explicitly shown in \Tab{tab:results}. This leads to an increase in the error estimate of the mass of the Milky Way. 

For reference, we also show the masses found by \cite{2018A&A...616L...9M,2019arXiv190102016D,2020arXiv200616283K}. We note there is a small difference in convention, where \cite{2018A&A...616L...9M} and \cite{2020arXiv200616283K} define the escape velocity cutoff as $3r_{340}$ and not $2 r_{200}$ as we do, although we do not expect more than a few percent difference due to this effect. We find that our analysis consistently shows a lower mass of the Milky Way than previous studies, although it is within a single standard deviation of the result from \cite{2019arXiv190102016D}. Note that we include the result from the retrograde data for completeness, and to show that even the non-convergent result for $\vesc$ still translates to a somewhat consistent total mass.

As can be seen in \Fig{fig:mass_concentration}, the circular velocity measurement plays an important role in constraining the mass of the Milky Way. For example, the current uncertainties on both the circular and escape velocities introduce a $\sim 20\%$ error on the mass estimate. Dropping the error on the circular velocity from 10 km/s to 5 km/s would reduce this error down to $\sim 15\%$. The choice of baryonic model also affects the estimate of the total mass of the Milky Way. More explicitly, if we were to adopt the baryonic model used in \cite{2014A&A...562A..91P} (which is based on \cite{2008ApJ...684.1143X}), we find $M_{200, \rm{tot}} = 6.4^{+1.3}_{-0.9} \times 10^{11}M_{\odot}$, or a $\sim 10\%$ difference from the value we quote above. A better understanding of the baryonic model of the Milky Way as well as a more accurate measurement of the circular velocity are therefore important in improving the total Milky Way mass estimates. Similarly, using a contracted NFW profile due to the presence of baryons \citep{Schaller15} would also change the total mass estimate substantially; \cite{2014A&A...562A..91P} used both the regular and contracted NFW profiles and found a difference of $\sim 50\%$.

\section{Conclusions}
\label{sec:conclusions}

In this paper, we applied a new analysis pipeline for constraining $\vesc$ that accounts for the presence of kinematic substructure in the stellar speed distribution. Our work is motivated by the discovery of the \Gaia Sausage-Enceladus, as well as by the need to improve the robustness of $\vesc$ fits, which can be very sensitive to the definition of the ``tail'' of the velocity distribution. We introduce a forward model that allows for multiple power law components in the tail of the stellar velocity distribution, and showed that repeating the analysis for $\vmin$ and number of components allows us to robustly determine $\vesc$. Details of this pipeline and examples with mock datasets can be found in Paper I.

We found that at least two bound components is preferred in fits to the \Gaia data for stars with speeds $v > 325$ km/s. These components are suggestive of a Sausage-Enceladus component and a relaxed stellar halo component, but more study is needed to understand if the components truly have different physical origin or if a multi-component model simply provides a more flexible fitting framework. With this multi-component fit, our result for $\vesc$ is lower than previous measurements in the literature. On the other hand, previous works assumed a single power law component for the tail, and we have shown that assuming a single component where the data prefers more can lead to systematically higher $\vesc$ values. 

Using our results for $\vesc$, we determined the total mass of the Milky Way, finding a value of the concentration of $c_{200} = 13.8^{+6.0}_{-4.3}$ and a total mass of $M_{200} = 7.0^{+1.9}_{-1.2} \times 10^{11} M_{\odot}$. Our result for the total mass is lower than those of previous studies relying on the escape velocity to obtain the mass of the Galaxy. However, they are more consistent with methods based on the distribution function of globular clusters (e.g. \cite{2019ApJ...875..159E}) and matching satellites of the Milky Way with their simulation counterparts (e.g. \cite{2018ApJ...857...78P}). \cite{2020SCPMA..6309801W} provides a review of all these methods, which shows a large scatter in the mass estimates ranging from $\sim (0.5 - 2) \times 10^{12} M_{\odot}$.  Our method provides the most robust measurement relying on the escape velocity at the location of the Sun. 

Along with a better understanding of the baryonic components of the Milky Way, the dark matter profile, and the local circular velocity, other effects are important to evaluate in order to improve Milky Way mass estimates. In particular, many of the existing estimates assume a relaxed equilibrated halo between the location of the Sun and the edge of the Galaxy.
The presence of satellites, streams, and the evidence of the active merger history of the Milky Way would suggest otherwise. It is therefore crucial to pair our pipeline, applied at different distances from the Galactic center, with a better understanding of the halo at larger radii. A combined approach will help build a complete and coherent picture of the Milky Way potential, and constrain the shape of the dark matter halo.

\section*{Acknowledgements}

We are grateful to I. Moult for early discussions and collaboration on the project, and to M. Lisanti for helpful feedback.
We would also like to thank L. Anderson, A. Bonaca, G. Collin, A. Deason, P. Hopkins, A. Ji, and J. Johnson for helpful conversations.  

This work was performed in part at Aspen Center for Physics, which is supported by National Science Foundation grant PHY-1607611.
This research used resources of the National Energy Research Scientific Computing Center (NERSC), a U.S. Department of Energy Office of Science User Facility operated under Contract No. DE-AC02-05CH11231. 
LN is supported by the DOE under Award Number DESC0011632, the Sherman Fairchild fellowship, the University of California Presidential fellowship, and the fellowship of theoretical astrophysics at Carnegie Observatories.
TL is supported by an Alfred P. Sloan Research Fellowship and Department of Energy (DOE) grant DE-SC0019195.

This work has made use of data from the European Space Agency (ESA) mission
{\it Gaia} (\url{https://www.cosmos.esa.int/gaia}), processed by the {\it Gaia}
Data Processing and Analysis Consortium (DPAC,
\url{https://www.cosmos.esa.int/web/gaia/dpac/consortium}). Funding for the DPAC
has been provided by national institutions, in particular the institutions
participating in the {\it Gaia} Multilateral Agreement.

\def\bibsection{} 
\bibliographystyle{aasjournal}
\bibliography{v_escape}

\pagebreak
\clearpage
\appendix

\setcounter{equation}{0}
\setcounter{figure}{0}
\setcounter{table}{0}
\setcounter{section}{0}
\makeatletter
\renewcommand{\theequation}{S\arabic{equation}}
\renewcommand{\thefigure}{S\arabic{figure}}
\renewcommand{\theHfigure}{S\arabic{figure}}
\renewcommand{\thetable}{S\arabic{table}}

\section{Data selection \label{app:datasel}}

\begin{figure*}[h] 
   \centering
	\includegraphics[width=0.7\textwidth]{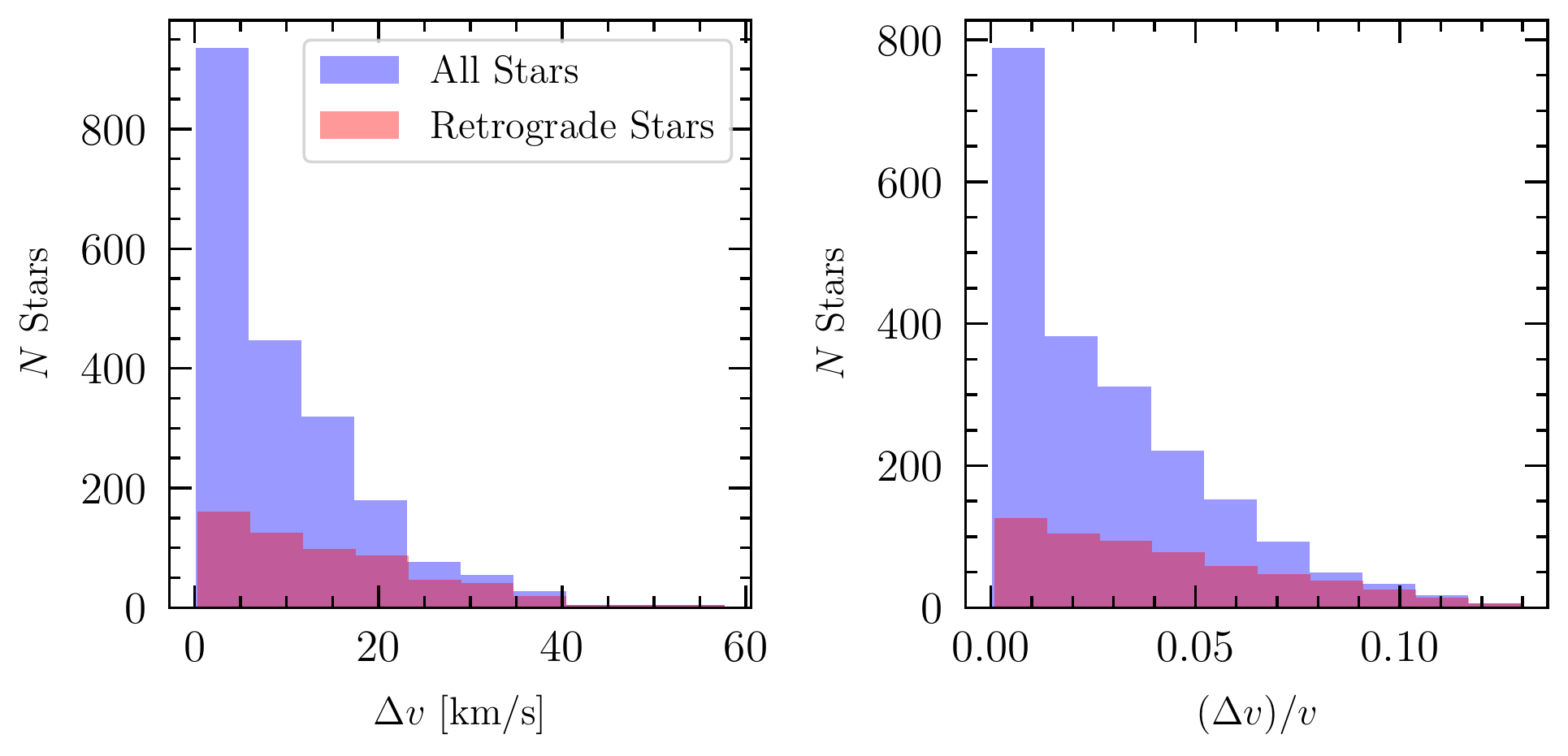} 
   \caption{The top panel shows the distribution of speed errors in the entire \Gaia sample satisfying distance and parallax error cuts, while the bottom panel shows the distribution of fractional speed error.    }
   \label{fig:data_cuts}
\end{figure*}

For the data sample in the main text, we apply Galactocentric distance cuts as well as parallax error cuts. For the remaining stars,  we show the distribution of the errors on the measured speeds in Fig.~\ref{fig:data_cuts}.  The cut placed in the main text of $\Delta v/v < 0.05$ only removes a small fraction of the entire data sample, while the typical errors in the retrograde sample are larger. We account for the error in individual stars in the likelihood, as discussed in \cite{methodology}.

\newpage

\FloatBarrier
\section{Corner plots for analyses in main text \label{app:corner} }

\subsection{Corner plots for $\vmin = 300$ km/s}

\begin{figure*}[h] 
   \centering
	\includegraphics[width=0.55\textwidth]{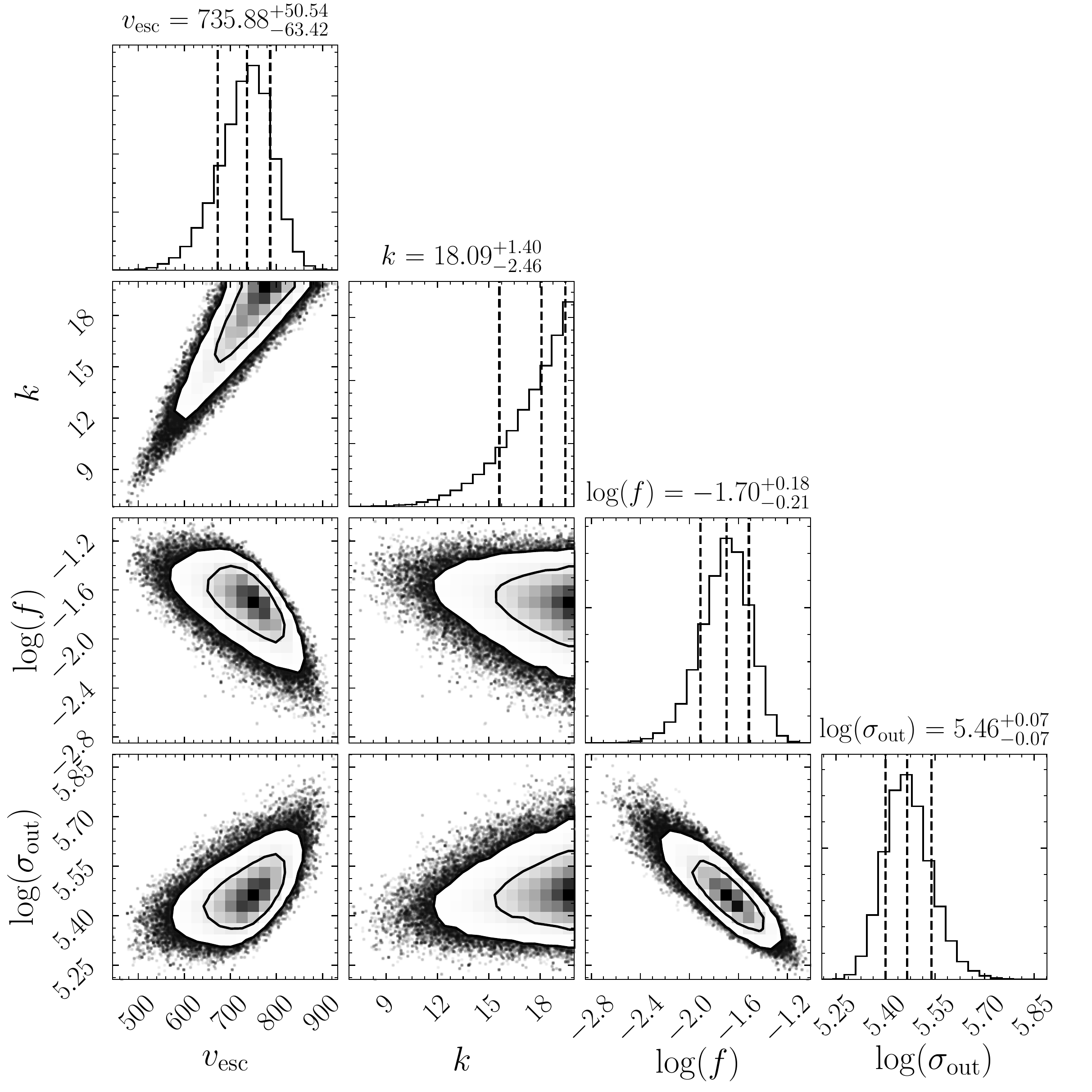} 
   \caption{Corner plot for the single component fit of the full \Gaia data, with $\vmin = 300$ km/s, and a cap of 5\% on the measured fractional error of the speeds. The contours correspond to 68\% and 95\% containment. }
   \label{fig:corner_one}
\end{figure*}

\begin{figure*}[h] 
   \centering
	\includegraphics[width=0.75\textwidth]{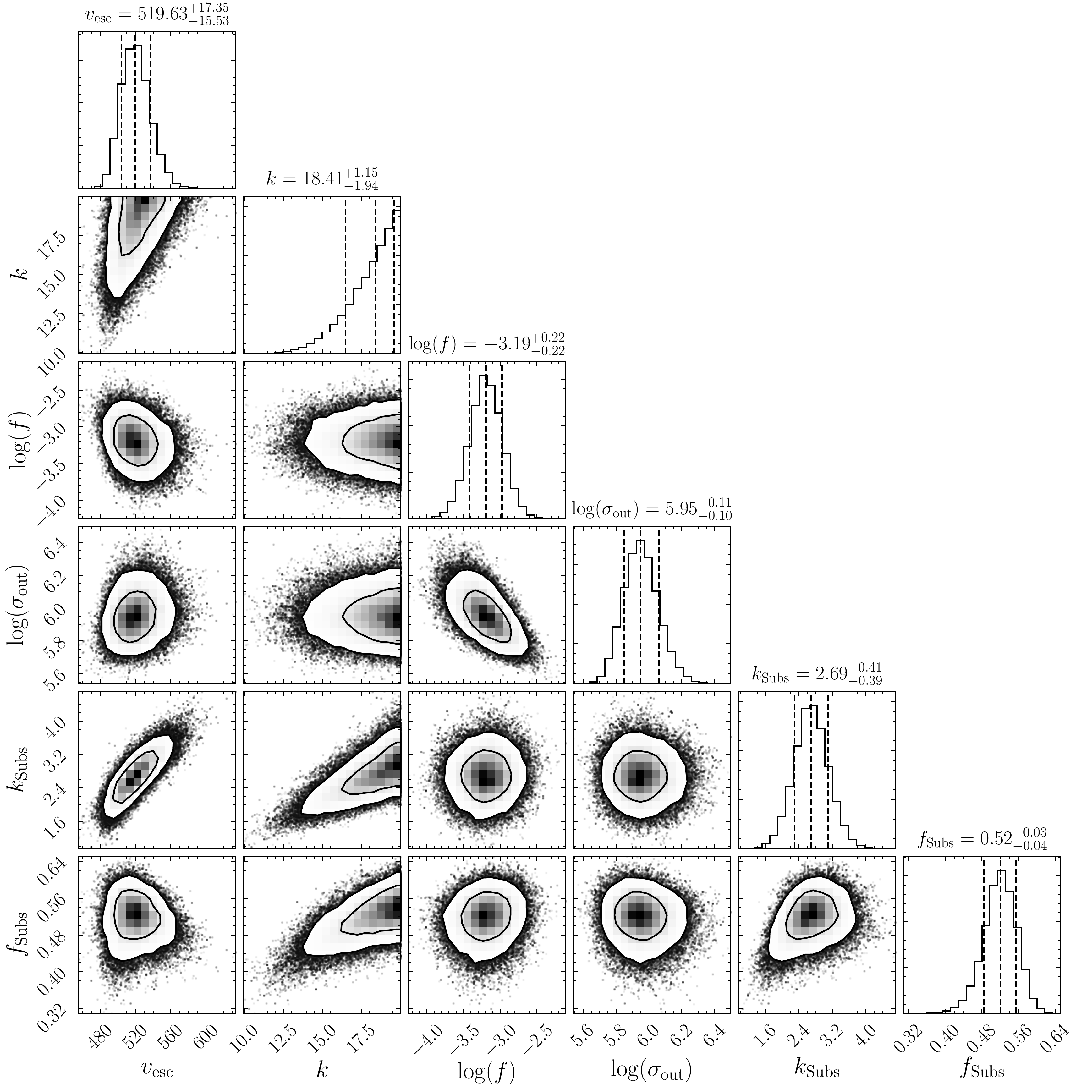} 
   \caption{Corner plot for the two component fit of the full \Gaia data, with $\vmin = 300$ km/s, and a cap of 5\% on the measured fractional error of the speeds. The contours correspond to 68\% and 95\% containment.}
   \label{fig:corner_two}
\end{figure*}

\begin{figure*}[h] 
   \centering
	\includegraphics[width=0.95\textwidth]{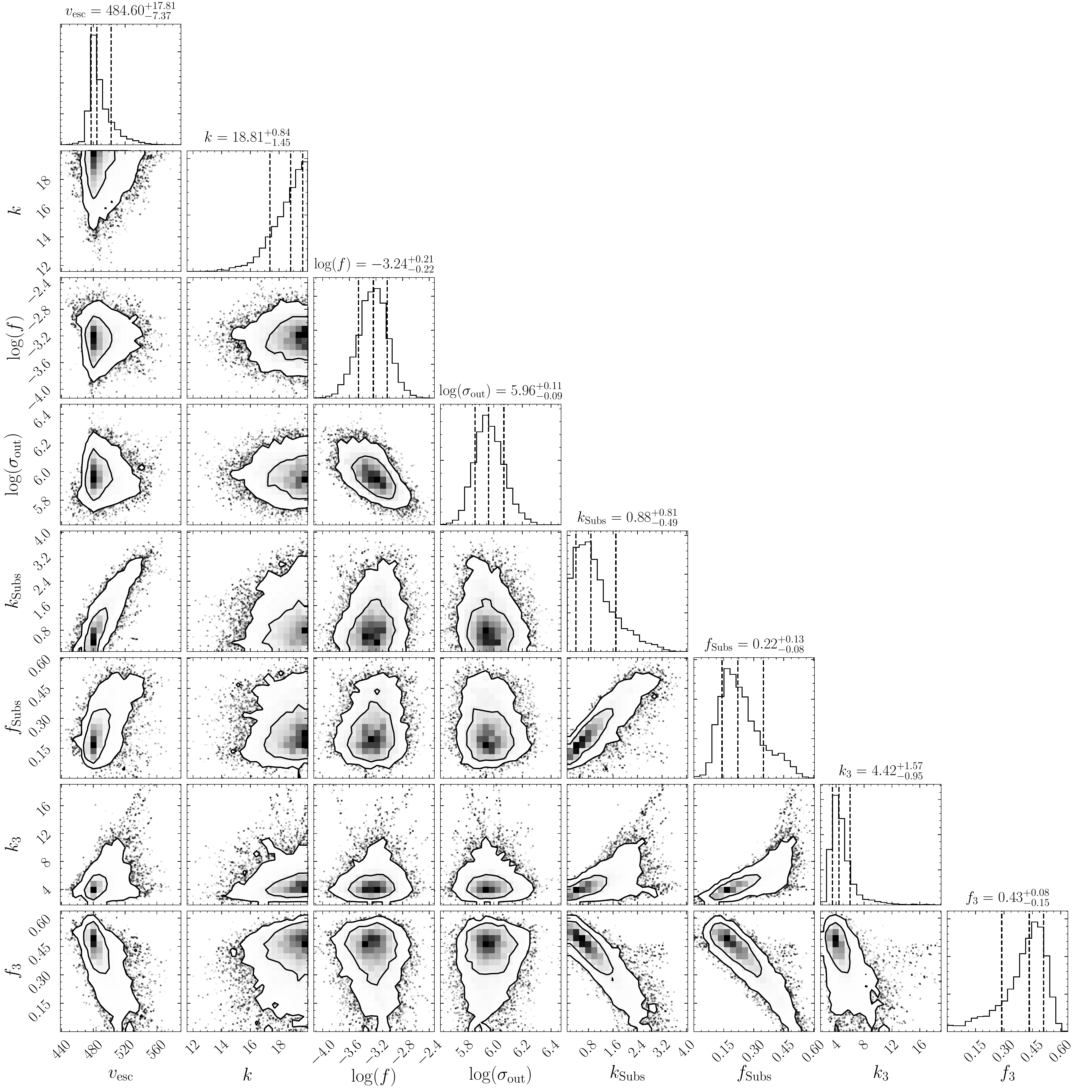} 
   \caption{Corner plot for the three component fit of the full \Gaia data, with $\vmin = 300$ km/s, and a cap of 5\% on the measured fractional error of the speeds.The contours correspond to 68\% and 95\% containment. }
   \label{fig:corner_three}
\end{figure*}

\FloatBarrier
\subsection{Corner plot for $\vmin = 350$ km/s and 2-component fit}

\begin{figure*}[h] 
   \centering
	\includegraphics[width=0.75\textwidth]{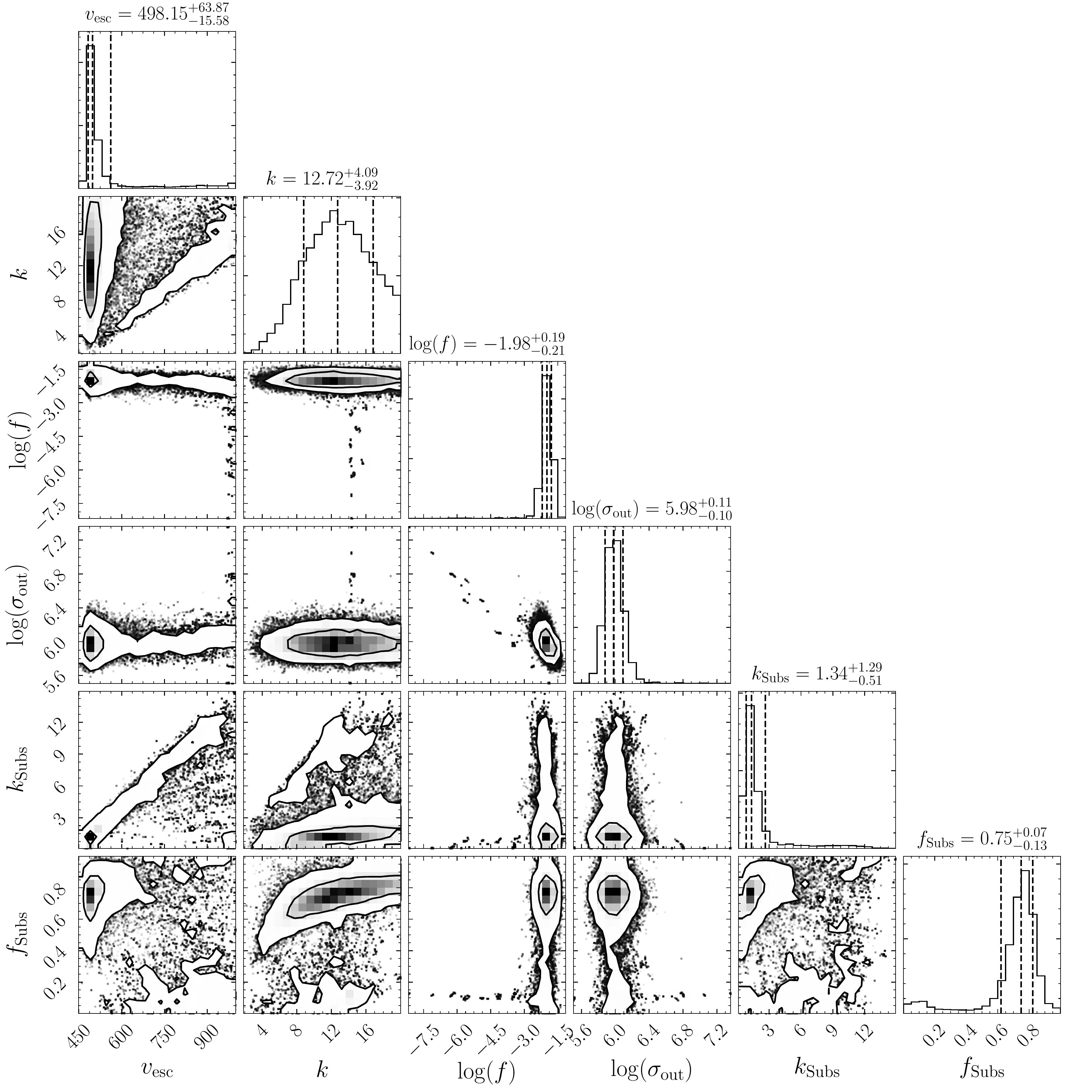} 
   \caption{Corner plot for the two component fit of the full \Gaia data, with $\vmin = 350$ km/s, and a cap of 5\% on the measured fractional error of the speeds. The contours correspond to 68\% and 95\% containment.}
   \label{fig:corner_two_350}
\end{figure*}

\clearpage

\FloatBarrier
\subsection{Corner plot for $\vmin = 375$ km/s and 1-component fit}

\begin{figure*}[h] 
   \centering
	\includegraphics[width=0.55\textwidth]{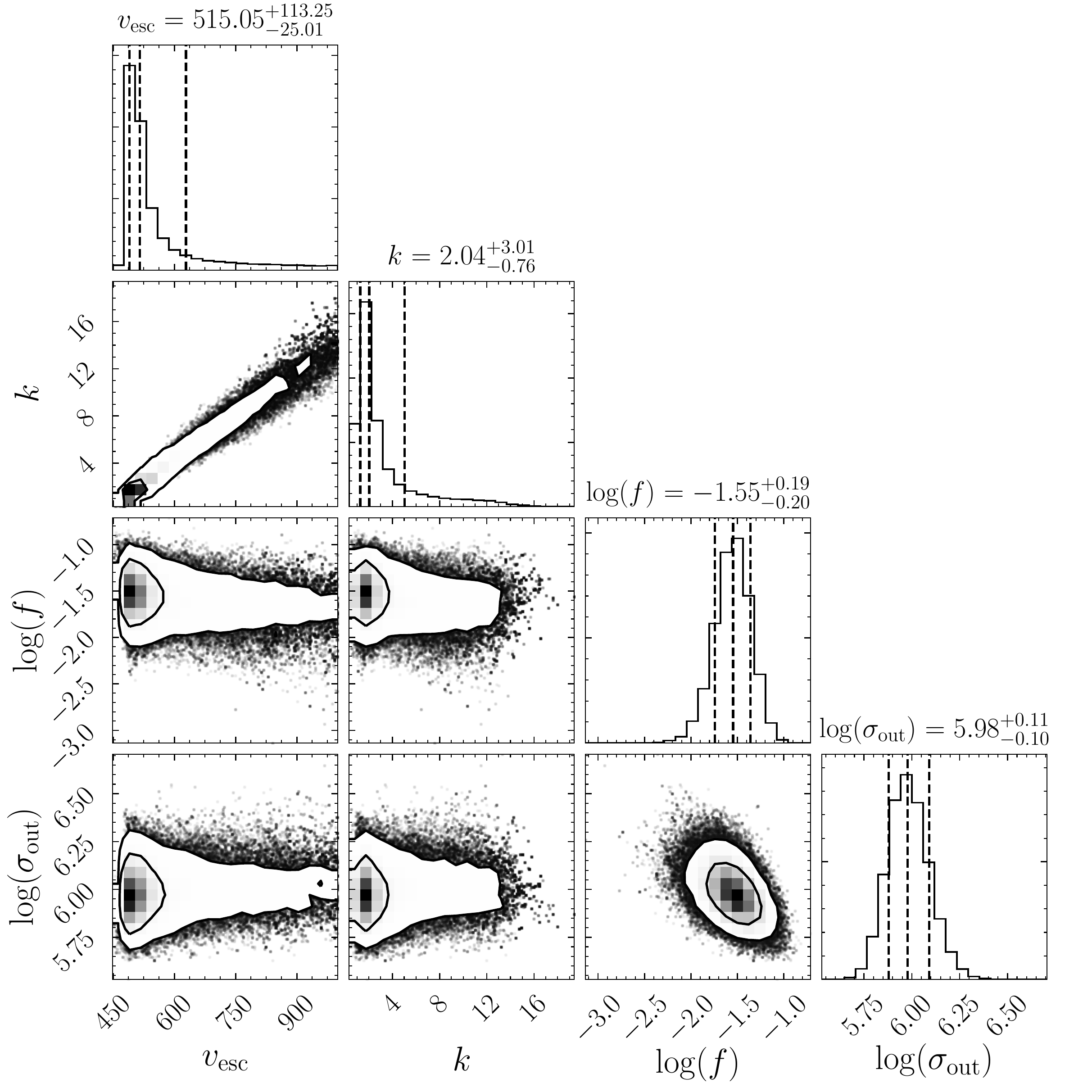} 
   \caption{Corner plot for the single component fit of the full \Gaia data, with $\vmin = 375$ km/s, and a cap of 5\% on the measured fractional error of the speeds. The contours correspond to 68\% and 95\% containment.}
   \label{fig:corner_one_375}
\end{figure*}

\clearpage

\FloatBarrier
\section{Analysis with Retrograde Data \label{app:retrograde}}

\begin{figure*}[h] 
   \centering
	\includegraphics[width=0.85\textwidth]{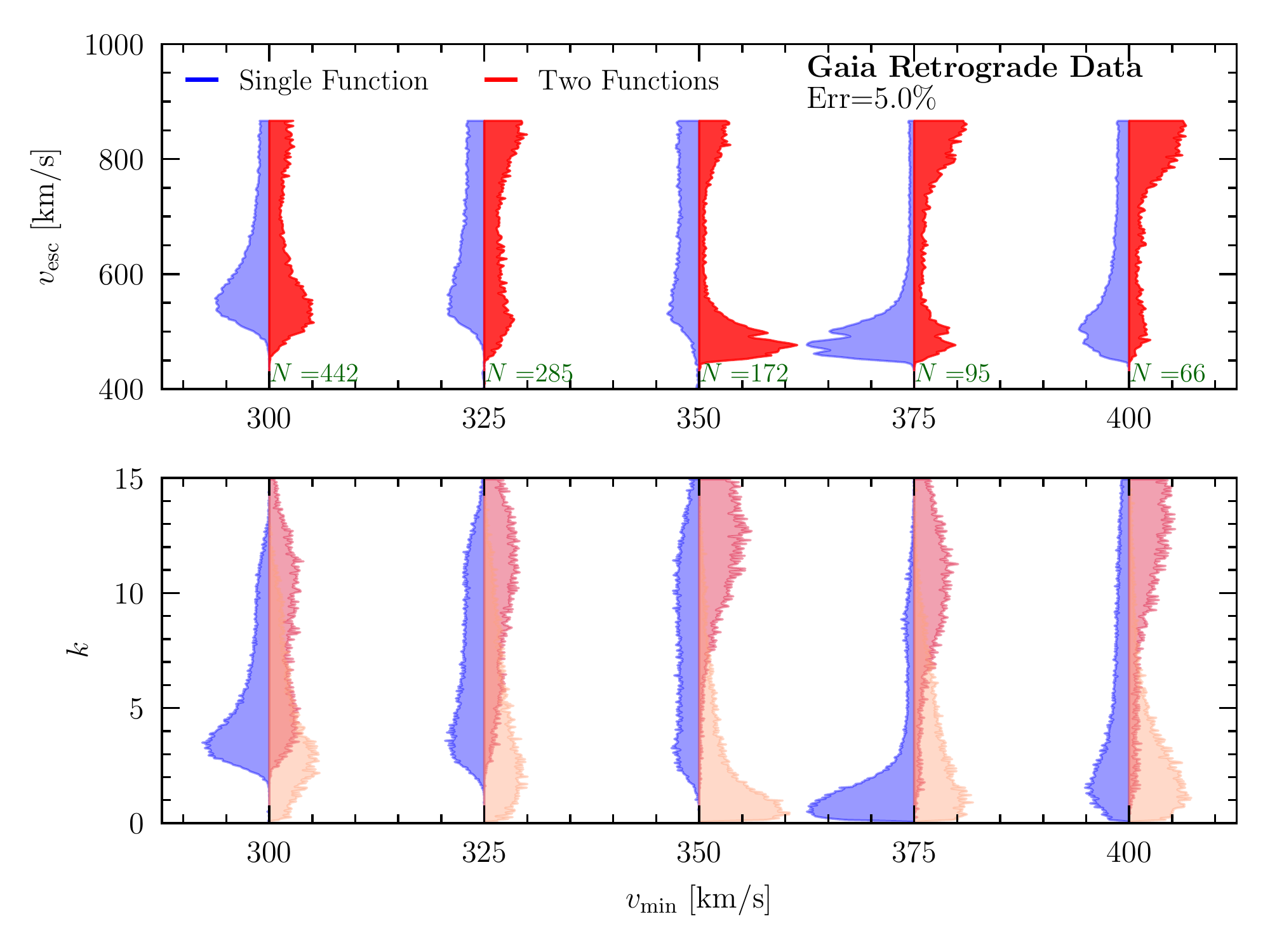} 
   \caption{From fitting the retrograde data only. Posteriors in escape velocity and $k$ for different values of $\vmin$, with single-component or two-component fits.}
   \label{fig:violin_retrograde}
\end{figure*}

In the main text, we found robust and self-consistent results for $\vesc$ when fitting the \Gaia dataset (satisfying some quality cuts). We are not able to find robust results when fitting retrograde stars only. In \Fig{fig:violin_retrograde}, we see that the $\vesc$ posteriors are non-convergent in all cases, and there is no clear trend in the results with $\vmin$.  As described in the main text and can be seen in \Fig{fig:retro}, there are far fewer retrograde stars with speeds above 500 km/s, and the outlier population is not as well constrained, leading to outlier confusion and the double-peaked posteriors.

\end{document}